\documentclass[preprint,pra,10pt,showpacs]{revtex4}
\usepackage{graphicx,amsmath,amssymb}

\newcommand{\beq}{\begin{equation}}
\newcommand{\eeq}{\end{equation}}
\newcommand{\beqa}{\begin{eqnarray}}
\newcommand{\eeqa}{\end{eqnarray}}
\newcommand{\ket}[1]{| #1 \rangle}
\newcommand{\bra}[1]{\langle #1 |}

\begin{document}


\title{Entanglement tensor for a general pure multipartite
quantum state}

\author{Hoshang Heydari}
\email{hoshang@imit.kth.se} \homepage{http://www.ele.kth.se/QEO/}
\affiliation{Department of Microelectronics and Information
Technology, Royal Institute of Technology (KTH), Electrum 229,
SE-164 40 Kista, Sweden}

\author{Gunnar Bj\"{o}rk}
\affiliation{Department of Microelectronics and Information
Technology, Royal Institute of Technology (KTH), Electrum 229,
SE-164 40 Kista, Sweden}

\date{\today}

\begin{abstract}
We propose an entanglement tensor to compute the entanglement of a
general pure multipartite quantum state. We compare the ensuing
tensor with the concurrence for bipartite state and apply the
tensor measure to some interesting examples of entangled
three-qubit and four-qubit states. It is shown that in defining
the degree of entanglement of a multi-partite state, one needs to
make assumptions about the willingness of the parties to
cooperate. We also discuss the degree of entanglement of the
multi-qubit $\ket{W_{M}}$-states.
\end{abstract}

\pacs{03.67.Mn, 42.50.Dv, 42.50.Hz, 42.65.Ky}

\maketitle

\section{Introduction}
Quantum theory is a fundamental theory that can describe the
subatomic world with a fascinating accuracy. Since 1935, quantum
entanglement \cite{Sch35,EPR35} has been central for the
understanding of the foundations of quantum theory. Besides, its
fundamental interest, entanglement has become an essential
resource for quantum communication applications created in recent
years, which have potential applications such as quantum
cryptography \cite{Bennett84,Ekert91} and quantum teleportation
\cite{Bennett93}.
 One  widely used measure of entanglement for
 a pair of qubits is  the concurrence, that gives an analytic formula for
the entanglement of formation
\cite{Bennett96,Wootters98,Wootters00}. In recent years, there
have been  proposals to generalize this measure to general
bipartite states, e.g., Uhlmann \cite{Uhlmann00} has generalized
the concept of concurrence by considering arbitrary conjugation,
then Audenaert, Verstraete, and De Moor \cite{Audenaert} in the
spirit of Uhlmann's work, generalized the measure  by defining a
concurrence vector for pure states. Another generalization of
concurrence has been done by Rungta \emph{et al.} \cite{Rungta01}
based on the idea of a super operator called universal state
inversion. Moreover, Gerjuoy \cite{Gerjuoy}, and Albeverio and Fei
\cite{Albeverio}, gave an explicit expression of generalized
concurrence in terms of the coefficients of a general, pure,
bipartite state. It is therefore interesting to be able to
generalized this measure from bipartite to multipartite systems
\cite{Bhaktavatsala,Akhtarshenas}. Quantifying entanglement of
multipartite states
\cite{Lewen00,Vedral97,Werner89,Plenio00,Hor00,Acin01,Soto02,Bennett96a,Dur99,ECKERT02,Dur00,Eisert01,Verst03},
is  complicated task. In \cite{Hosh3,Hosh2,Hosh1} we proposed a
measure of entanglement for a general pure multipartite state. In
this paper we will take one step further in  generalizing of
concurrence to multipartite states by giving an explicit formula
for this measure and verifying the well establish result for
bipartite states. Then, we give some results related to the
simplest example of a multipartite state, namely a tripartite
state. Moreover, we give some examples of four-partite states in
such way as to show the idea behind our measure. Finally we will
prove D\"{u}r {\em et al's} conjecture about the general
multi-qubit $\ket{W_{M}}$ state. The concurrence of a two qubit,
bipartite state is defined as
  $\mathcal{C}(\Psi)=|\langle\Psi\ket{\widetilde{\Psi}}|$,
  where  the tilde represents the "spin-flip" operation
  $\ket{\widetilde{\Psi}}=\sigma_{y}\otimes
  \sigma_{y}\ket{\Psi^{*}}$, $\ket{\Psi^{*}}=\sum^{2}_{l,k=1}\alpha^{*}_{k,l}\ket{k,l}$ is the complex
  conjugate of $\ket{\Psi}=\sum^{2}_{l,k=1}\alpha_{k,l}\ket{k,l}$, and $\sigma_{y}=\left(%
\begin{array}{cc}
  0 & -i \\
  i & 0 \\
\end{array}%
\right)$ is a Pauli spin-flip operator. It was noted by Peres
\cite{Peres} that for a separable state, $\ket{\widetilde{\Psi}}$
is orthogonal to $\ket{\Psi}$, whereas for any Bell-state, the
states are parallel. The concurrence can also be written as
follows
  \begin{eqnarray}
\mathcal{C}(\Psi)&=&|\langle\Psi\ket{\widetilde{\Psi}}|\\\nonumber
  &=&|\mathrm{Tr}(\sigma_{y}\otimes
  \sigma_{y}\ket{\Psi}\bra{\Psi^{*}})|.
\end{eqnarray}
In the following section we will introduce an entanglement tensor,
of a form similar to concurrence. For a bipartite qubit state, our
measure  coincide with the concurrence, or by the generalization
of concurrence developed \cite{Gerjuoy,Albeverio}.  In contrast,
e.g., the bipartite entanglement of a tripartite state depends on
what action the third party takes. It is possible to find states
where the third party can vary the remaining two parties' state
continuously between a separable state and a maximally entangled
state. Only the $M$-partite entanglement of an $M$-partite state
can be defined unambiguously.

\section{Entanglement tensor for general pure  multipartite
quantum state}

In this section, we will give an expression for the entanglement
of a general pure multipartite state. The derivation of the
measure is tedious, and follows almost exactly that of our measure
based on the density matrix of a pure state \cite{Hosh3}.
Therefore, it will not be repeated here. It suffices to point out
that the mathematical derivation of the measure is based on the
relative-phase correlations between a quantum system's various
sub-systems.

Let
\beq\ket{\Psi}=\sum^{N_{1}}_{k_{1}=1}\sum^{N_{2}}_{k_{2}=1}\cdots\sum^{N_{m}}_{k_{m}=1}
\alpha_{k_{1},k_{2},\ldots,k_{m}}
\ket{k_{1},k_{2},\ldots,k_{m}},\label{eq: state} \eeq be a general
pure state defined on the Hilbert space
$\mathcal{H}_{\mathcal{Q}_{1}}\otimes\mathcal{H}_{\mathcal{Q}_{2}}\otimes\cdots\otimes
  \mathcal{H}_{\mathcal{Q}_{m}}$.

We can also introduce projection probabilities by projecting the
state $\ket{\Psi}$ onto the basis states in one or more of the
subspaces $\mathcal{H}_{\mathcal{Q}_{r}}$ and computing the norm
of the projection. E.g., reducing the $j$:th subspace, we get the
probabilities \beq p_{k_j} = \bra{\Psi}k_j\rangle \langle k_j
\ket{\Psi} . \eeq In the same vein the projection probabilities if
we project onto the $j$:th and $r$:th subspace, we get \beq
p_{k_j,k_r} = \bra{\Psi}k_j,k_r\rangle \langle k_j,k_r \ket{\Psi}
.\eeq

We also need an index permutation operator $\mathrm{P}_{j}$
operating on the state coefficient product
$\alpha_{k_{1},k_{2},\ldots,k_{j},\ldots,k_{m}}\alpha_{l_{1},l_{2},
\ldots,l_{j},\ldots,l_{m}}$ as follows
\begin{eqnarray}
&&\nonumber
\mathrm{P}_{j}(\alpha_{k_{1},k_{2},\ldots,k_{j},\ldots,k_{m}}\alpha_{l_{1},l_{2},
\ldots,l_{j},\ldots,l_{m}})\\\nonumber &&=
\alpha_{k_{1},k_{2},\ldots,k_{j},\ldots,k_{m}}\alpha_{l_{1},l_{2},
\ldots,l_{j},\ldots,l_{m}}\\\nonumber &&
-\alpha_{k_{1},k_{2},\ldots,l_{j},\ldots,k_{m}}\alpha_{l_{1},l_{2},
\ldots,k_{j},\ldots,l_{m}}.
\end{eqnarray}
In an $M$-partite state, there are many ways to share
entanglement. There are e.g. $M(M-1)/2$ different kinds of of
bipartite entanglement, entanglement that can be shared between
parties 1 and 2, 1 and 3, et.c. until parties $M-1$ and $M$. In
general, there are \beq \left ( \begin{array}{c} M \\ D
\end{array} \right ) = \frac{M !}{D! (M-D)!} \eeq different kinds
of $D$-partite entanglement in an $M$-partite state, where $M \geq
D$. Each of these components have an associated entanglement
tensor coefficient. Using our permutation operator above, we can
define a $D$-partite tensor coefficient $c_{r, \ldots,z}$,
containing information about the entanglement between the $D$
parties $r, \ldots,z$, where parties $r, \ldots,z$ can be chosen
any way among the $M$, as \begin{eqnarray} c_{r, \ldots,z} & = &
\left ( \mathcal{N}_{D} \sum^{N_{1}}_{k_{1}=1} \cdots
\sum^{N_{r-1}}_{k_{r-1}=1} \sum^{N_{z+1}}_{k_{z+1}=1} \cdots
\sum^{N_{M}}_{k_{M}=1} (p_{k_1,\ldots,k_{r-1},k_{z+1}, \ldots ,k_M})^{-1} \nonumber \right . \\
& & \left . \sum^{N_{r}-1}_{l_{r}>k_{r}}\sum^{N_{r}-1}_{k_{r}=0}
 \cdots
\sum^{N_{z}-1}_{l_{z}>k_{z}}\sum^{N_{z}-1}_{k_{z}=0}
|\mathrm{P}_{r+1}|\mathrm{P}_{r+2}\cdots|\mathrm{P}_{z}
(\alpha_{k_{1},k_{2},\ldots,k_{M-1},k_{M}}\alpha_{k_{1}\ldots,k_{r-1},l_r
\ldots,l_{z},k_{z+1},k_{M}}) |^{2}\cdots ||  \right )^{1/2}
\label{eq:generic tensor component}\end{eqnarray}

Assume that we have a state where subsystem $j$ is separable from
all other subsystems. In such a case, it holds that
$\alpha_{k_{1},k_{2},\ldots,k_{j},\ldots,k_{m}}\alpha_{l_{1},l_{2},\ldots,l_{j},\ldots,l_{m}}=
\alpha_{k_{1},k_{2},\ldots,l_{j},\ldots,k_{m}}\alpha_{l_{1},l_{2},\ldots,k_{j},\ldots,l_{m}}
$. That is, every entanglement tensor component involving the
entanglement between subsystem $j$ and any other subsystem(s) is
identically zero. Hence, separability of any subsystem can
directly be detected by looking at all entanglement tensor
components associated with a certain subsystem. Note that one
needs to look through all different kinds of entanglement
(bipartite, tripartite, etc.) to ensure separability. We also see
that the expression for $c_{r, \ldots,z}$ is independent of local
phase-transformations, e.g. transformations of the type \beq
\sum_{k_j=1}^{N_J} e^{i \phi_{k_j}} \ket{k_j}\bra{k_j} ,\eeq where
$\phi_{k_j}$ are real numbers, because such a transformation will
result in the same change of phase in the factors
$\alpha_{k_{1},k_{2},\ldots,k_{j},\ldots,k_{m}}\alpha_{l_{1},l_{2},\ldots,l_{j},\ldots,l_{m}}$
and
$\alpha_{k_{1},k_{2},\ldots,l_{j},\ldots,k_{m}}\alpha_{l_{1},l_{2},\ldots,k_{j},\ldots,l_{m}}
$.

\section{Concurrence for  bipartite quantum
states} \label{sec:bipartite} As we have already mentioned, there
has been considerable progress to generalize concurrence for
bipartite states in arbitrary dimensions
\cite{Gerjuoy,Albeverio,Akhtarshenas,Hosh3,Bhaktavatsala}. As our
first example, we show that our entanglement tensor component
(there is only one component for a bipartite state) coincide with
the well established formula for the generalized concurrence of a
bipartite state. Let
$\ket{\Psi}=\sum^{N_{1}-1}_{k_{1}=0}\sum^{N_{2}-1}_{k_{2}=0}\alpha_{k_{1},k_{2}}\ket{k_{1},k_{2}}$
be a general pure state defined on a bipartite  Hilbert space
$\mathcal{H}_{\mathcal{Q}_{1}}\otimes\mathcal{H}_{\mathcal{Q}_{2}}$.
Then, the bipartite entanglement tensor component of the state is
given by \beq c_{12}=\left ( \mathcal{N}_{2}
\sum^{N_{1}}_{l_{1}>k_{1}}\sum^{N_{1}}_{k_{1}=1}
\sum^{N_{2}}_{l_{2}>k_{2}}\sum^{N_{2}}_{k_{2}=1}
\left|\alpha_{k_{1},k_{2}}\alpha_{l_{1},l_{2}}-\alpha_{k_{1},l_{2}}\alpha_{l_{1},k_{2}}\right|^{2}
\right )^{\frac{1}{2}} , \eeq where, if we choose the
normalization constant $\mathcal{N}_{2}=4$, that is, a
normalization constant based on setting the entanglement of an
EPR-pair to unity, we get identically the concurrence of the state
\cite{Gerjuoy,Albeverio}. In particular, for a pair of qubits
\cite{Wootters00}, we have \beq c_{12} = \mathcal{N}_{2}
|\alpha_{1,1}\alpha_{2,2}-\alpha_{1,2}\alpha_{2,1}|.\eeq The
component is independent of any unitary operations, local to
subsystems 1 and 2.

\section{Entanglement of tripartite quantum states}
\label{sec:tripartite}

The first step towards the more complex states goes through the
tripartite state, which is the ``simplest'' state that can be
called a multipartite state. Let
$\ket{\Psi}=\sum^{N_{1}-1}_{k_{1}=0}\sum^{N_{2}-1}_{k_{2}=0}
\sum^{N_{3}-1}_{k_{3}=0}\alpha_{k_{1},k_{2},k_{3}}\ket{k_{1},k_{2},k_{3}}$
be a general pure state. This state has three bipartite
entanglement tensor components and one tripartite tensor
component. They are: \beq c_{12}= \left
 (\mathcal{N}_{2} \sum^{N_{3}}_{k_{3}=1} p_{k_3}^{-1} \sum^{N_{1}}_{l_{1}>k_{1}}\sum^{N_{1}}_{k_{1}=1}\sum^{N_{2}}_{l_{2}>k_{2}}\sum^{N_{2}}_{k_{2}=1}
| \alpha_{k_{1},k_{2},k_{3}}\alpha_{l_{1},l_{2},k_{3}}
-\alpha_{k_{1},l_{2},k_{3}}\alpha_{l_{1},k_{2},k_{3}} |^{2} \right
)^{1/2} , \eeq \beq c_{13}=\left
 (\mathcal{N}_{2} \sum^{N_{2}}_{k_{2}=1} p_{k_2}^{-1}\sum^{N_{1}}_{l_{1}>k_{1}}\sum^{N_{1}}_{k_{1}=1}
\sum^{N_{3}}_{l_{3}>k_{3}}\sum^{N_{3}}_{k_{3}=1} |
\alpha_{k_{1},k_{2},k_{3}}\alpha_{l_{1},k_{2},l_{3}}
-\alpha_{k_{1},k_{2},l_{3}}\alpha_{l_{1},k_{2},k_{3}} |^{2} \right
)^{1/2} , \eeq \beq c_{23}=\left
 (\mathcal{N}_{2} \sum^{N_1}_{k_{1}=1} p_{k_1}^{-1} \sum^{N_{2}}_{l_{2}>k_{2}}\sum^{N_{2}}_{k_{2}=1}
\sum^{N_{3}}_{l_{3}>k_{3}}\sum^{N_{3}}_{k_{3}=1}  |
\alpha_{k_{1},k_{2},k_{3}}\alpha_{k_{1},l_{2},l_{3}}
-\alpha_{k_{1},k_{2},l_{3}}\alpha_{k_{1},l_{2},k_{3}} |^{2} \right
)^{1/2} , \eeq and \beq c_{123}= \left (
\mathcal{N}_{3}\sum^{N_{1}-1}_{l_{1}>k_{1}}\sum^{N_{1}-1}_{k_{1}=0}
\sum^{N_{2}-1}_{l_{2}>k_{2}}\sum^{N_{2}-1}_{k_{2}=0}
\sum^{N_{3}-1}_{l_{3}>k_{3}}\sum^{N_{3}-1}_{k_{3}=0} \{||
\alpha_{k_{1},k_{2},k_{3}}\alpha_{l_{1},l_{2},l_{3}}-
\alpha_{k_{1},k_{2},l_{3}}\alpha_{l_{1},l_{2},k_{3}}|^{2}-
|\alpha_{k_{1},l_{2},k_{3}}\alpha_{l_{1},k_{2},l_{3}}
-\alpha_{k_{1},l_{2},l_{3}}\alpha_{l_{1},k_{2},k_{3}} |^{2}|
\}\right )^{1/2} .\eeq In this case, the bipartite tensor
components are, in general, not independent of local unitary
transformations (except for local phase shifts). Instead, the
entanglement of multipartite states depends, in general, both on
local operations and on whether or not the parties choose to
cooperate. That is, the local operations one party chooses to
perform on his subsystem, and the extent to which he chooses to
communicate his result, determines the entanglement of the
remaining state. A necessary requirement for an entanglement
measure is its monotonicity under local operations and classical
communication. The measure should not increase under such
transformations. However, if one makes a local measurement on a
multipartite state, both the amount and the form of the
entanglement may be changed. Our entanglement tensor component as
given by (\ref{eq:generic tensor component}), is not monotonic
under local transformations. Hence, the entanglement of a state
must be defined as the supremum of (\ref{eq:generic tensor
component}) under all unitary transformations. However, there is
an intrinsic problem with such an optimization. It is well known
that, e.g., tripartite entanglement may be transformed into
bipartite entanglement and vice versa. Neither transformation is
reversible. One can get a maximum of one EPR-state per initial GHZ
state. At the same time, in the limit of many EPR-states, we can
only obtain 2 GHZ-states from 3 EPR-states \cite{Bennett96a}. The
optimal conversion rates between most tripartite and
higher-partite states are still unknown. Before such conversion
rates are known, (and a classification of the irreversible sets of
states is done \cite{Dur00,Pan}) it is not possible to give
appropriate weights to the tripartite, fourpartite, etc. tensor
components. This implies that until then, it is only possible to
find the supremum of our entanglement measure for each kind of
entanglement separately \cite{Hosh3}. This precludes proper
entanglement quantification of, e.g. the state \beq (\ket{1,1,0} +
\ket{1,0,1} + \ket{0,1,1} + \ket{1,0,0})/2, \eeq a state that
contains both bipartite and tripartite entanglement and that
cannot be converted by invertible local operators neither to a
W-state nor to a GHZ-state \cite{Pan}.

We shall see below that if an $M$-partite state has $D$-partite
entanglement, where $M>D$, and we assume that the subsystems are
labelled such that we want to quantify the entanglement between
parties $1, \ldots, D$, then the supremum of our measure assumes
that parties $D+1, \ldots , M$ cooperate with the parties $1,
\ldots, D$.

Let us first study the W-state $\ket{W_{3}}$ that is given by \beq
\ket{W_{3}} =
\frac{1}{\sqrt{3}}(\ket{1,0,0}+\ket{0,1,0}+\ket{0,0,1}) . \eeq The
tripartite entanglement tensor component $c_{123}$ of this state
is zero, and it can be shown that it remains zero under all local
transformations. Each of the state's three bipartite tensor
components' supremal values are equal to \beq c_{12} =c_{13}
=c_{23} =\sqrt{\frac{\mathcal{N}_{2}}{6}} .\eeq The state is known
for its robustness under loss of one qubit. If any of the three
qubits is traced out, the ensuing mixed two-qubit state has the
same average entanglement as the original pure three-qubit state.

Next, consider the GHZ-state \beq \ket{{\rm GHZ}} =
\frac{1}{\sqrt{2}}(\ket{0,0,0}+\ket{1,1,1}). \eeq The bipartite
tensor components of the state in this basis are all zero, whereas
the tripartite tensor component attains its maximal value $c_{123}
= \sqrt{\mathcal{N}_{3}}/2$. Now assume that a Hadamard
transformation is made on the leftmost qubit. The ensuing state
becomes \beq
\frac{1}{2}(\ket{0,0,0}+\ket{1,0,0}+\ket{0,1,1}-\ket{1,1,1}). \eeq
The state in this basis has $c_{123}=c_{12}=c_{13}=0$ and
$c_{23}=\sqrt{\mathcal{N}_{2}}/2$. The component $c_{23}$ reaches
its supremum in this basis. This result can easily be interpreted.
If the leftmost qubit is measured in the computational basis, the
results zero and unity will occur with equal probability, 1/2. If
one obtains the result zero, the remaining state will be in the
EPR-state $(\ket{0,0}+\ket{1,1})/\sqrt{2}$. If one obtains the
result unity, then the he remaining state will be in the EPR-state
$(\ket{0,0}-\ket{1,1})/\sqrt{2}$, orthogonal to the one above.
However, if the measurement result is communicated to the parties
holding the remaining two qubits, either party can convert one of
the EPR-states to the other using local operations (a local phase
shift). Therefore, irrespective of the measurement result, the
remaining state can be made to be a deterministic EPR-state, and
this is what our result predicts. If, on the other hand, the
measurement result is not communicated, then the ensuing
bipartite, qubit mixed state is separable.

In order to use the state's symmetry to the fullest, now suppose
that all three qubits of the GHZ-state are Hadamard-transformed.
The ensuing state is \beq \ket{\bar{W}} =
\frac{1}{2}(\ket{0,0,0}+\ket{0,1,1}+\ket{1,0,1}+\ket{1,1,0}). \eeq
This state has $c_{123}=0$ and $c_{12}=c_{13}=c_{23}=
\sqrt{\mathcal{N}_{2}}/2$, and this is the basis in which all
three components $c_{12}$, $c_{13}$, and $c_{23}$ simultaneously
attain their suprema. In this case, measurement of the value of
any of the three qubits and subsequent communication of the result
will enable the parties holding the remaining two qubits to
transform their state into a deterministic EPR-state. We see that
the entanglement tensor components give the entanglement of the
corresponding state, provided that the parties cooperate. In this
case, the entanglement of each bipartite subsystem is equal to
that of a EPR-state. Hence, the average entanglement of the
$\ket{\bar{W}}$ state is higher than that of the W-state, a state
that is sometimes referred to as the most biparte entangled
tripartite state. The latter statement is true if one assumes that
one of the qubits is simply discarded, corresponding to a
trace-operation. If, however, the parties chose to cooperate, the
state $\ket{\bar{W}}$ has a higher average bipartite entanglement.

In earlier papers \cite{Hosh1,Hosh2,Hosh3}, we have defined the
entanglement in a way that can be interpreted as a tensor norm.
Such a crude measure has some merit. However, as only one number
is obtained, a large norm does not signal whether or not the state
is highly entangled (a GHZ-state being the simple example), or if
the state is not highly entangled, but has entanglement ``all over
the place'' (such as a W-state). Giving all the entanglement
tensor components rather than the norm of the tensor of course
gives more information about the particular type of entanglement
of a state.


\subsection{Entanglement of four-partite quantum states}
\label{sec:four-partite}

As a first example of four qubit state, let us consider the state
\beq \ket{\Psi}
=\frac{1}{2}(\ket{0,1,1,0}+\ket{1,0,0,1}+\ket{0,1,1,1}+\ket{1,0,0,0})
. \eeq The state has no four-partite entanglement and in the given
computational basis, it has no bipartite entanglement. The
tripartite entanglement tensor components $c_{124}$,$c_{134}$,and
$c_{234}$ are all zero, while, inserting the state's expansion
coefficients in (\ref{eq:generic tensor component}) we have
\begin{eqnarray}
c_{123}&=&(2\mathcal{N}_{3}[|\alpha_{1,2,2,1}\alpha_{2,1,1,1}|^{2}
+|\alpha_{1,2,2,2}\alpha_{2,1,1,2}|^{2}])^{\frac{1}{2}}\\\nonumber
&=&(2\mathcal{N}_{3}[\frac{1}{16}
+\frac{1}{16}])^{\frac{1}{2}}=\sqrt{\frac{\mathcal{N}_{3}}{4}}.
\end{eqnarray}
It is quite clear that this is the supremal value of this
tripartite tensor component. The result can most easily be checked
by writing the state
\begin{eqnarray}
\ket{\Psi}\nonumber
&=&\frac{1}{\sqrt{2}}(\ket{0,1,1}+\ket{1,0,0})\otimes
\frac{1}{\sqrt{2}}(\ket{0}+\ket{1}) . \label{eq:visible product}
\end{eqnarray}
In this case, any local action on the rightmost qubit will not
change the state's its entanglement. However, as shown in the
previous section, local actions on the remaining three qubits may
transform the tripartite entanglement to various degrees of
bipartite entanglement.

The state \beq \ket{\psi} = \frac{1}{2\sqrt{2}}\left([\ket{0,0} +
\ket{1,1}]\otimes[\ket{0,1} + \ket{1,0}] + [\ket{0,1} +
\ket{1,0}]\otimes[\ket{0,0} + \ket{1,1}]\right) ,\label{eq:nested
entanglement}\eeq is an example of a state that has nested
entanglement. That is, the state is a (bipartite) entangled state
of (bipartite) entangled states. Computing the entanglement for
this state in the given basis, we find that the state has no
four-partite entanglement, no tripartite entanglement, whereas all
six bipartite entanglement tensor components are equal to
$\sqrt{\mathcal{N}_{2}}/2$, indicating EPR-type entanglement.
Again, cooperation between the parties is needed to exploit this
entanglement. However, this state has the feature that we also can
see it as a bipartite $\mathcal{H}_4 \otimes \mathcal{H}_4$ state,
if each of two parties have access to two of the qubits. The
bipartite $\mathcal{H}_4 \otimes \mathcal{H}_4$ entanglement of
the state can also be obtained by the expression (\ref{eq:generic
tensor component}). In this particular case we get the supremal
value $\sqrt{\mathcal{N}_{2}}/2$.

It is obvious that, in general, a state's entanglement depend on
the chosen Hilbert space factorization of the state.
Operationally, this can be stated that the entanglement of the
state depends on how the state's subsystems are shared among the
parties because this division defines what operations are
considered to be local. This is why, in this paper, we have made a
distinction between subsystems and parties.

 As our last example of a four-qubit state, consider the four
 qubit W-state
\beq \ket{W_{4}} =
\frac{1}{2}(\ket{0001}+\ket{0010}+\ket{1000}+\ket{0100}). \eeq
Quite expectedly, the state has no four-partite, nor any
tripartite entanglement. The supremal values of the six bipartite
entanglement tensor components are all equal to $
\sqrt{\mathcal{N}_{2}/8}$. The state is robust to the loss of any
two qubits, and a rather obvious analysis show that the parties
need not cooperate to get this result. Note, that the state
$\ket{\psi}$ in Eq. (\ref{eq:nested entanglement}), above, give a
substantially higher average bipartite entanglement, but only if
the parties cooperate.

\section{Entanglement of multi-qubit W-states}
\label{sec:m-partite}

As a very last example, we would like to show the applicability of
formula (\ref{eq:generic tensor component}) even on generic
classes of multipartite states. A simple case of a multipartite
state is the generalization of $\ket{W_{3}}$ and $\ket{W_{4}}$ to
$\ket{W_{M}}$, where $M$ signifies $M$ qubits. This state can
symbolically be written as
\begin{equation}
\ket{W_{M}}=\frac{1}{\sqrt{M}}\ket{M-1,1},
\end{equation}
where $\ket{M-1,1}$ denotes the totally symmetric superposition
state including $M-1$ zeros and 1 one. The entanglement of this
state is, again, very robust against particle losses, i.e., the
state $\ket{W_{M}}$ remains entangled even if any $M-2$ parties
discards, or loses, the information about their particle.

In a paper by D\"{u}r, Vidal, and Cirac, \cite{Dur00}, it is
conjectured that the average value of the square of the
concurrence for $\ket{W_{M}}$ is given by
\begin{equation}
\frac{2}{M(M-1)}\sum_{k}\sum_{k\not=l}\mathcal{C}^{2}_{k,l}(W_{M})=\frac{4}{M^{2}}
\end{equation}
The expression for the entanglement tensor, Eq. (\ref{eq:generic
tensor component}), gives the result that all tensor components
are equal to zero, except for the bipartite components that all
simultaneously can have the supremal values $\sqrt{\mathcal{N}_2/2
M}$. As discussed in Sec. \ref{sec:bipartite}, we should set
$\mathcal{N}_2=4$ to make our tensor components equal to unity for
an EPR-state. Doing so, we obtain the value $2/M$ for all of the
tensor components squared. That is, the average of the components
squared is also $2/M$. The interpretation of this result is rather
simple. If all but two qubits of the state is lost, we stand a
$2/M$ chance of having an EPR-pair and a $(M-2)/M$ chance of
having the state $\ket{0,0}$. From a large number $N$ of
$\ket{W_{M}}$-states, we can hence statistically obtain $2 N/M$
pure EPR states. However, as demonstrated for tri- and
four-partite systems, this is not the highest achievable average
(bipartite) entanglement for a $M$-partite state. This (the
concurrence squared) is instead $\mathcal{N}_2/4$, or unity if
$\mathcal{N}_2$ is set to four. This result assumes that the $M-2$
qubits are not lost but measured, and that the parties cooperate.


\section{Discussion and conclusion}

In conclusion we have proposed an explicit formula for an
entanglement tensor of a general, pure, multipartite quantum
state. To demonstrate the nature of the measure, and some of the
aspects involved in entanglement classification such as
cooperation, we have given some example for bipartite, tripartite,
four-partite, and $M$-partite states. In Sec. \ref{sec:m-partite}
we confirm the conjecture by D\"{u}r {\it et al.} about the
concurrence of multi-qubit $\ket{W_{M}}$-states. However, we note
that a higher value of the average concurrence of a state is
possible, provided that the parties cooperate. That is, that the
unused qubits are not simply ignored.

\begin{acknowledgments}
This work was supported by the Swedish Research Council (VR) and
the Swedish Foundation for Strategic Research (SSF).
\end{acknowledgments}

\end{document}